\documentclass[12pt]{article}
\usepackage{epsfig,amssymb,amsmath,psfrag,subfigure,rotate,color,wasysym}

\allowdisplaybreaks
\newcommand{\insertfig}[2]{\includegraphics[width=#1cm]{#2}}

\def\XXint#1#2#3{{\setbox0=\hbox{$#1{#2#3}{\int}$ }
\vcenter{\hbox{$#2#3$ }}\kern-.6\wd0}}

\def \be  {\begin{equation}}
\def \ee  {\end{equation}}
\def \ba  {\begin{eqnarray}}
\def \ea  {\end{eqnarray}}
\def \baa {\begin{eqnarray*}}
\def \eaa {\end{eqnarray*}}
\def \lab #1 {\label{#1}}

\newcommand\re[1]{(\ref{#1})}
\def\d{\hbox{{d}\kern-.20em\hbox{l}}}

\def \matrix #1 {\left(\begin{array}{cc} #1 \end{array}\right)}

\def \tr {\mathop{\rm tr}\nolimits}

\newcommand \vev [1] {\langle{#1}\rangle}
\newcommand \VEV [1] {\left\langle{#1}\right\rangle}
\newcommand \ket [1] {|{#1}\rangle}

\newcommand{\bit}[1]{\mbox{\boldmath$#1$}}
\def\1{\hbox{{1}\kern-.25em\hbox{l}}}

\newcommand{\ft}[2]{{\textstyle\frac{#1}{#2}}}
\begin{document}

\begin{titlepage}

\thispagestyle{empty}

\vspace*{3cm}

\centerline{\large \bf Separation of Variables for a flux tube with an end}
\vspace*{1cm}

\centerline{\sc A.V. Belitsky}

\vspace{10mm}

\centerline{\it Department of Physics, Arizona State University}
\centerline{\it Tempe, AZ 85287-1504, USA}

\vspace{2cm}

\centerline{\bf Abstract}

\vspace{5mm}

We consider a partial light-cone limit of a correlation function of the stress-tensor multiplet and identify an integrable structure emerging at one loop order
of perturbation theory. It corresponds to a noncompact open spin chain with one boundary being recoil-less while the other one fully dynamical. We solve 
the system by means of techniques of the Baxter operator and Separation of Variables. The eigenvalues of the separated variables define rapidities of 
excitations propagating on the color flux tube and encode their factorizable dynamics in the presence of a dynamical boundary. 

\end{titlepage}

\setcounter{footnote} 0

\newpage

\section{Introduction}

As we now know from the gauge/string correspondence \cite{Maldacena:1997re}, planar Yang-Mills theories are, in fact, string theories in a disguise. This allows one to map
complicated dynamics occurring in real space-time to the one of the world-sheet. The latter is in turn amenable to treatments devised for two-dimensional systems,
which are much simpler indeed. Sometimes they can even be integrable and, therefore, corresponding physical observables found exactly for any value of the
't Hooft coupling $a = g^2_{\scriptstyle\rm YM} N_c/(2 \pi)^2$.

To date, the best studied example of this kind is the maximally supersymmetric Yang-Mills theory. The latter is a superconformal interacting theory. Any two-point correlations 
functions of composite operators are known exactly in this model since the corresponding spectral problem for anomalous dimensions was solved thanks to its integrability,
see \cite{Beisert:2010jr} and references cited therein. Recently, higher point correlation functions were addressed within a framework of the so-called hexagon expansion
\cite{Basso:2015zoa}, which relies on a tessellation of the two-dimensional world-sheet defining the correlation function in the dual string description in terms of certain form 
factors which can be found exactly from a set of axioms valid nonperturbatively. 

A multiple pair-wise light-cone limit of the aforementioned correlators gives access to vacuum expectation values of Wilson loops on null polygonal contours in Minkowski space-time
\cite{Alday:2010zy}. These in turn were found to be in a dual pair with scattering amplitudes \cite{Alday:2009yn} of properly regularized $\mathcal{N} = 4$ super-Yang-Mills 
theory. It is the Wilson loop side that provides a viable language for two-dimensional description: by singling out two nonadjacent sides of the loop, we can think of them as a
``quark-antiquark'' pair propagating with a speed of light and sourcing out a flux tube between them, which from the point of view of holography looks like a one-dimensional 
string projected on the boundary. The string sweeps a two-dimensional world-sheet which turns out to be integrable as well \cite{Basso:2013vsa}.

In this paper we study a somewhat hybrid of a function, which is obtained from multi-point correlation functions by taking a partial light-cone limit. The advantage of this 
kinematics is that it allows one to probe boundary interactions of the flux-tube attached to a dynamical rather than recoil-less ``quark''. We find that, again, particle-like 
excitations propagating on top of the flux with an end have diffractionless scattering and can be solved exactly. Presently, we analyze physical observables to one-loop
order and uncover that the physics is encoded in a one-dimensional model of noncompact Heisenberg spins living on an semi-infinite interval. We solve this model within the 
framework of the Baxter operator and Separation of Variables (SoV). We identify the eigenvalues of a complete set of charges with the momentum injected into the recoiled 
boundary and rapidities of flux-tube excitations.

Our subsequent consideration is organized as follows. In the next section, we introduce correlation functions with all operators placed on a two-dimensional Minkowski plane
and take their partial pairwise light-cone limit reducing our analysis to a study of correlation functions of certain light-ray operators. We address a particular class of one-loop
corrections in Section \ref{OneLoopSection} that is driven by a non-local renormalization group evolution of these operators, which is then brought into the form of a Hamiltonian 
system for a collection of noncompact spins. The Hilbert space of the model and an inner product defined on it are introduced in Section \ref{ModelSection}. Next, we give 
a lightning overview of the formalism of factorized $R$-matrix in Section \ref{RmatrixSection}, which is used to build the Baxter operator in Section \ref{BaxterQsection}. We find a 
finite-difference relation, known as the Baxter equation, which it obeys in Section \ref{SectionBaxterEq}. Remarkably, multiparticle wave functions for this magnet can be found 
explicitly in an integral form on certain multi-variable two-dimensional graphs as addressed in Section \ref{EigenfunctionsSection}. In fact, these are nothing else as the wave 
functions of an off-diagonal element of the monodromy matrix analyzed more than a decade ago in Ref.\ \cite{Derkachov:2003qb}. Finally, we conclude. Throughout our analysis, 
we heavily rely on a Feynman diagram approach to verify and prove various statements. In spite of the fact that the rung moves in Feynman graphs had already appeared a 
dozen of times in the literature before, we will repeat them in the Appendix, along with a few of other ingredients, for integrity of our presentation.

\section{Partial light-like limit}

In this paper we are going to relax the strict pairwise light-like limit which led to the Wilson loop stretched on a null polygonal contour \cite{Alday:2010zy}.  To simplify our consideration, 
we will place all operators on a two-dimensional surface R$^{1,1}$. This is a special kinematics akin to the one discussed within the context of scattering amplitudes \cite{Alday:2009yn}. 
The first nontrivial Wilson loop in this kinematics was an octagon. To draw analogies to the consideration at hand, we will presently address this case as well. 

We start with an eight-point bosonic correlation function,
\begin{align}
G_8 = \VEV{\prod_{j = 1}^4 O (z_j) \bar{O} (z_{j+1})}
\, , 
\end{align}
where the operators sitting at the odd and even positions are specific components of a protected 1/2 BPS operator $O^{ABCD}$, built from six real scalars of the vector multiplet 
$\phi^{AB} = - \phi^{BA} = \varepsilon^{ABCD}  \bar\phi_{CD}$ (with $A,B = 1,2,3,4$), transforming in the ${\bf 20'}$ representation of the SU(4) R-symmetry group. Namely,
\begin{align}
O = \tr \varphi^{2}
\, , \qquad
\bar{O} = \tr \bar\varphi^{2} 
\, .
\end{align}
where $\varphi \equiv \phi^{12}$ and $\bar\varphi \equiv \phi^{34}$. The partial light-like limit we are currently considering involves sending all consecutive points to become light-like 
separated $z_{jj+1}^2 \equiv (z_j - z_{j+1})^2 \to 0$ except one, say, $z_{12}^2 \neq 0$,
\begin{align}
F_8
\equiv 
\lim_{\{z^2_{jj+1} \}\backslash z_{12}^2 \to 0} \left( G_8 /\prod_{j=2}^8 D^{\rm tree}_{j j+1} \right)= \tr \vev{ D_A (z_1, z_2) [z_2,z_3] \dots [z_8,z_1]}_A
\, .
\end{align}
Here we factored out a free scalar propagator $D^{\rm tree}_{jj+1} \equiv \vev{\varphi (z_j) \bar\varphi (z_{j+1})}|_{g_{\rm\scriptscriptstyle YM} = 0} = - 1/(4 \pi^2 z^2_{jj+1})$ with 
the remainder given by the product of the path-ordered exponents in the adjoint representation
\begin{align}
[z_j, z_{j+1}] = P \exp \left( \ft{i}{2} g_{\rm\scriptscriptstyle YM} \int_{z_j}^{z_{j+1}} dz^{\dot\alpha \alpha}  A_{\alpha\dot\alpha} (z) \right)
\, .
\end{align}
This phase is the only modification of the leading singularity an interacting particle propagator acquires compared to the free theory \cite{Gross:1971wn,Efremov:1978xm}. They are path 
integral averaged over SU$(N_c)$ gauge fluctuations $A_{\alpha\dot\alpha}$, $\vev{\dots}_A$ with the exact scalar propagator $D_A (z_1, z_2) = \vev{\varphi (z_1) \bar\varphi (z_2)}$ 
in the external field $A_{\alpha\dot\alpha}$. 

\begin{figure}[t]
\begin{center}
\mbox{
\begin{picture}(0,140)(275,0)
\put(0,-300){\insertfig{22}{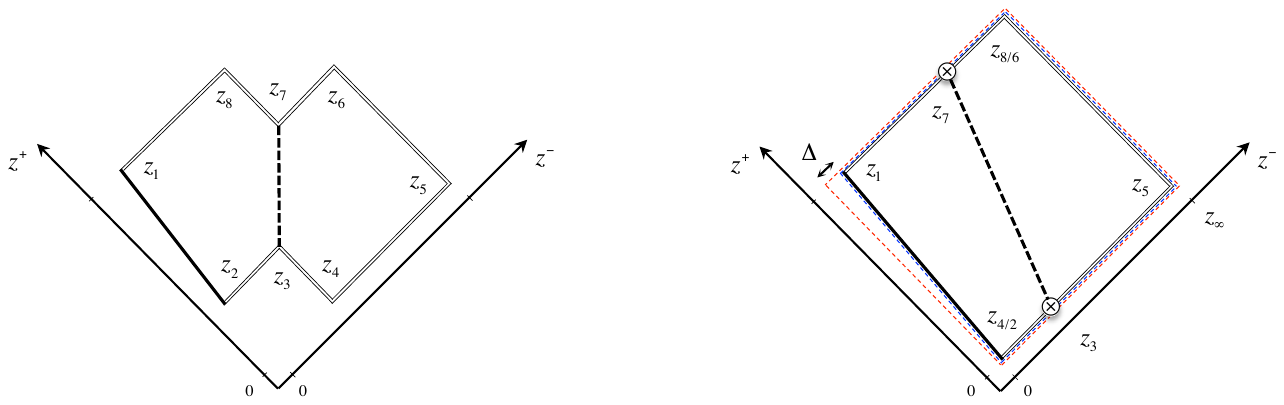}}
\end{picture}
}
\end{center}
\caption{ \label{2Doctagon} Two-dimensional octagonal correlation function.
}
\end{figure}

The above scalar operator $O^{ABCD}$ is a superconformal primary state of the $\mathcal{N} = 4$ stress-tensor multiplet, which contains among other states, 
the energy-momentum tensor of the theory. So considering the chiral superspace extension, we define echoing \cite{Eden:2011yp},
\begin{align}
\mathcal{G}_8 =  \VEV{\prod_{j = 1}^4 \mathcal{T} (Z_j)  \bar{\mathcal{T}}(Z_{j+1})}
\end{align}
with the  chiral stress tensor operator\footnote{Instead of using specific components, SU(4) covariance can be achieved by means of auxiliary harmonic variables.}
\begin{align}
\mathcal{T} = \tr W^{12} W^{12}
\, , \qquad
\bar{\mathcal{T}} = \tr W^{34} W^{34}
\end{align}
built from the superfield
\begin{align}
W^{AB} (Z) = \phi^{AB} (z) - i \theta^{\alpha [A} \psi_\alpha^{B]} (z)+ \dots
\, ,
\end{align}
that depends on the chiral space coordinate $Z = (z^{\alpha\dot\alpha}, \theta^{\alpha A})$ and contains the above ${\bf 20'}$ as its lowest component, i.e., when all Grassmann 
variables are set to zero, $\theta = 0$. The partial light-like limit now yields the result
\begin{align}
\mathcal{F}_8 
\equiv
\lim_{\{z^2_{jj+1} \}\backslash z_{12}^2 \to 0} \left( \mathcal{G}_8 /\prod_{j=2}^8 D^{\rm tree}_{j j+1} \right) = \tr \vev{ D_W (Z_1, Z_2) [[Z_2,Z_3]] \dots [[Z_8,Z_1]]}
\, , 
\end{align}
where the $[[Z_j,Z_k]]$ stands for a super-Wilson link
\begin{align}
[[Z_j,Z_k]]
=
P \, \exp
\left(
\ft{i}{2} g_{\rm\scriptscriptstyle YM} \int_{Z_j}^{Z_k} \left[ dz^{\dot\alpha\alpha} \mathcal{A}_{\alpha\dot\alpha} + 2 d \theta^{\alpha A} \mathcal{F}_{\alpha A} \right]
\right)
\, ,
\end{align}
determined by the bosonic $\mathcal{A}_{\alpha\dot\alpha} = {A}_{\alpha\dot\alpha} + O (\theta)$ and fermionic $\mathcal{F}_{\alpha A} = \ft{i}{2} \bar\phi_{AB} \theta^{\alpha B} + 
O (\theta^2)$ connections, and where $D_W (Z_1, Z_2)$ stands for the $W$-propagator. Obviously, setting all fermionic coordinates to zero, one gets back its bosonic counterpart, 
$\mathcal{F}_8 |_{\theta = 0} = F_8$. Introducing two-dimensional conjugate Weyl spinors $\ket{j}$ and $|j]$ for each light-cone distance $z_{jj+1} = \ket{j}[j|$, one defines
projected fermionic variables $\chi^A_j = \vev{j | \theta^A}$, which keep track of the quantum numbers of flux-tube excitations in the pentagon picture to scattering amplitudes 
\cite{Basso:2013vsa}. Below, we focus on a single Grassmann component of the supercorrelator, namely proportional to $\chi_2 \chi_3 \chi_5 \chi_6 \equiv \varepsilon_{ABCD} 
\chi^A_2 \chi^B_3 \chi^C_5 \chi^D_6$,
\begin{align}
\mathcal{F}_8 
=
\dots + \chi_2\chi_3 \chi_5\chi_6 F_{8;1} + \dots
\, .
\end{align}
At leading order in 't Hooft coupling, it is merely given by the product of two propagators as shown in the left panel of Fig.\ \ref{2Doctagon},
\begin{align}
\label{TreeOPE}
F_{8;1}^{\rm tree} = \frac{a}{4 \pi^2} \frac{1}{\vev{23} \vev{67} z_{37}^2 z_{12}^2}
\, ,
\end{align}
where in the two-dimensional kinematics $z_{jk}^2 = 2 z_{jk}^+ z_{jk}^-$ factorizes in terms of the light-cone coordinates $z^\pm$. Our choice is driven by its simplicity. In addition, when 
all distances are taken to the light cone, including $z_{12}^2$, i.e., $\Delta = 0$ in Fig.\ \ref{2Doctagon}, it goes into a NMHV amplitude induced by a flux-tube scalar exchange at tree 
level, see, e.g., Ref.\ \cite{Belitsky:2012mp}.

In this work, we focus on the interpretation of the above formula \re{TreeOPE} in terms of a light-cone operator product expansion. The operators in questions are of the form of
a $\Gamma$-shaped cusped Wilson line contour with a field $\varphi$ sourcing the chromoelectric field at the origin\footnote{In what follows, all coordinates will refer to the $z^-$
light ray and we will drop corresponding superscript designating this, unless it is ambiguous.},
\begin{align}
\mathcal{O}_{\Gamma} (0, z_1, z_2, \dots, z_\infty) = \varphi (0) [0,z_1] \varphi (z_1) [z_1, z_2] \dots W [z_\infty]
\, .
\end{align}
Here $[z_k, z_k]$ is a straight link long the $z^-$ direction while $W[z]$ starts at $z$ and runs along the $z^+$ axis as shown in the right panel of Fig.\ \ref{2Doctagon}. Then, Eq.\ 
\re{TreeOPE} is determined by the correlation function of two of these
\begin{align}
\label{2PointCF}
F_{8;1}^{\rm tree} = g^2_{\rm\scriptscriptstyle YM} \vev{\mathcal{O}_{\Gamma} (0, z_3, z_\infty) \bar{\mathcal{O}}_{\Gamma} (z_1, z_5, z_\infty) }|_{g_{\rm\scriptscriptstyle YM} = 0}
\, ,
\end{align}
where we denoted the right-most coordinate as $z_\infty$($= z_8$, for the octagon). Let us move on to the one-loop order next.

\section{One-loop renormalization}
\label{OneLoopSection}

Above, we were rather cavalier in our approach to the partial light-cone limit. While it was inconsequential at tree level, it needs to be properly addressed when quantum effects 
are accounted for as loop diagrams yield divergencies. Let us introduce, in the spirit of factorization theorems, a scale $\mu$ that will separate inverse distances involved 
and measure the deviation of an interval from the light ray. The light-cone limit then implies that $z_{jj+1}^2 \mu^2 \ll 1$, such the propagation of particles along corresponding 
sides is recoil-less since their virtuality $q^2 \sim z^{-2}_{jj+1} \gg \mu^{2}$. In other words, particles move very fast and observe their surroundings as a long wave-length 
external field, which does not distort their motion in an abrupt fashion. Their effect does not change the singularity structure of the free propagation \cite{Gross:1971wn,Efremov:1978xm}. 
The momentum of the gluon that travels on the $z^-$ interval is of order $1/z_\infty$ and it sets the scale of the soft radiation $\mu$. On the contrary, the particle propagating along the 
$z_{12}$ interval has the energy of order or less than $\mu$, $z_{12}^2 \mu^2 \sim 1$, and, therefore, gets recoiled. We will adopt the following nomenclature in what follows: we 
will call the Wilson loop links as the {\sl hard} boundary while the $z_{12}$-side as the {\sl soft} one.

Let us focus on the large-time evolution logarithms $\tau = \ft12 \ln z_{12}^+$ of the  correlation function \re{2PointCF}. It is plagued by collinear singularities. So a question
arises what ratio one has to form that makes the observable in question finite, on the one hand, while still staying sensitive to the boundary dynamics, on the other. These quantum
corrections can be encoded in terms of a light-ray Hamiltonian acting on the fields propagating in the exchange channel,
\begin{align}
\label{LRcorrfunc}
\vev{\mathcal{O}_{\Gamma} (0, z_3, z_\infty) \bar{\mathcal{O}}_{\Gamma} (z_1, z_5, z_\infty) }|_{g_{\rm\scriptscriptstyle YM} = 0} 
\to 
\mathcal{G}
\equiv
\vev{\mathcal{O}_{\Gamma} (0, z_3, z_\infty)  (1 + a \tau \mathcal{H}) \bar{\mathcal{O}}_{\Gamma} (z_1, z_5, z_\infty) }
\, ,
\end{align}
where
\begin{align}
\label{TotalHamiltonian}
\mathcal{H} = h_{01} + h_{1\infty} + h_{0 \infty}
\, ,
\end{align}
with the SL(2)-invariant pairwise Hamiltonian $h_{jk}$ acting on a field $X$ with conformal spin $s$
\begin{align}
h_{jk} X(z_j) X(z_k)
=
\int_0^1 \frac{d \alpha}{\bar\alpha} \left[ \alpha^{2 s - 1} X( \alpha z_j + \bar\alpha z_k) X(z_k) + \alpha^{2 s - 1} X(z_j) X( \alpha z_k + \bar \alpha z_j) - 2  X(z_j) X(z_k) \right]
\, .
\end{align}
The action of $\mathcal{H}$ on the tree-level function yields (here we set $s = \ft12$)
\begin{align}
\mathcal{H} F_{8;1}^{\rm tree} 
=
\left(
\ln \frac{z_{37} z_1}{z_{31} z_7}
+
\ln \frac{\varepsilon z_{37}}{z_\infty^2}
+
\ln \frac{\varepsilon z_{1}}{z_\infty^2}
\right)
F_{8;1}^{\rm tree} \, ,
\end{align}
corresponding to the three terms in Eq.\ \re{TotalHamiltonian}. Above, we regularized the intrinsic collinear divergence by deviating the hard boundary off the light cone $\varepsilon 
\equiv z_{8/6} - z_5$ and took the light-ray operator $z_\infty$ to be very long, i.e., larger than any other light-cone distances involved in accord with the flux-tube interpretation
\cite{Alday:2007mf}.

If we were to adopt the same reasoning as in the formation of the ratio function used in the amplitude framework, see, e.g., Ref.\ \cite{Basso:2013vsa}, we would normalize the above 
light-cone correlation function \re{LRcorrfunc} to the one without the insertion of the flux-tube excitation propagating from $z_3$ to $z_7$, see Fig.\ \ref{2Doctagon}. In that case, both 
boundaries were recoil-less and thus not dynamical and resulted into the subtraction of $2 h_{0 \infty}$ from Eq.\ \re{TotalHamiltonian}. In the present case, the left boundary is soft 
and gets recoiled. We therefore, would only like to get rid of the interaction term between the soft and hard boundaries (on the back face of the correlator), without over-subtraction of 
the physics of recoil (on its front). This implies that we have to take the square root of the light-cone correlator without the flux-tube insertions $G_\Box = \mathcal{G}|_{\rm no\ flux-tube\ 
insertions}$, rather than its whole and, thus, subtract only $h_{0 \infty}$. The function $G_\Box$ corresponds to the correlator of light-cone operators with the (blue) dashed contour in 
the right panel of Fig.\ \ref{2Doctagon}. However, we immediately observe that collinear logarithms are not completely cancelled. To accomplish this, we have to additionally divide the 
correlator by the square root of the rectangular Wilson $W_\Box$ in the fundamental representation over the (red) square contour in Fig.\ \ref{2Doctagon}. The ratio function then to study 
is\footnote{At higher orders of perturbation theory, the front and back faces of the ratio $\mathcal{R}$ start interacting and their factorization is lost.}
\begin{align}
\label{SubtractedR}
\mathcal{R} = \frac{\mathcal{G}}{\sqrt{G_{\Box} W_{\Box}}}
\, .
\end{align}
The large-time one-loop corrections to the resulting observable effectively emerge from the following Hamiltonian
\begin{align}
\mathcal{H}_{\mathcal R} = h_{01} + \widetilde{h}_{1\infty} 
\, , 
\end{align}
with the same SL(2) invariant Hamiltonian between the light boundary and the flux-tube excitation, but a modified one for the interaction between the flux-tube excitation and
the hard Wilson line boundary,
\begin{align}
\widetilde{h}_{j\infty} X (z_j) W (z_\infty)
=
\int_0^1 \frac{d \alpha}{\bar\alpha} \left[ \alpha^{2 s - 1} X( \alpha z_j + \bar\alpha z_\infty) W (z_\infty) - X(z_j) W (z_\infty) \right]
-
\ln (\mu z_{j\infty})
X(z_j) W (z_\infty) 
\, .
\end{align}
where the factorization scale $\mu \equiv 1/z_\infty$, introduced for obvious dimensional reasons, separates the soft and hard gluon radiation. This Hamiltonian is obviously not SL(2) invariant 
and was considered before within the context of heavy-light hadrons \cite{Lange:2003ff,Braun:2003wx} and $\mathcal{N} = 4$ SYM scattering amplitudes \cite{Belitsky:2011nn}. These 
Hamiltonians can be re-written in terms of generators of the collinear conformal algebra as (see, e.g., \cite{Belitsky:2014rba,Braun:2014owa}),
\begin{align}
\label{PairwiseHamiltonians}
h_{jk}
=
2 \psi (1) - 2 \psi (J_{jk})
\, , \qquad
\widetilde{h}_{j \infty} = \psi (1) - \ln \left( \mu S^+_j \right)
\, , 
\end{align}
where the arguments of the digamma functions are given in terms of the pairwise Casimir $J_{jk} (J_{jk} - 1) = (\bit{S}_j + \bit{S}_k)^2$ and components of the sl(2) generators to be 
introduced in the next section.

It is now straightforward to place any number of flux-tube excitations on the $z^-$ light rays on the top and bottom sides of the square. In the multicolor limit, their interaction Hamiltonian is 
merely given by the sum of pairwise nearest-neighbor interactions such that for $N$ of them, we have
\begin{align}
\label{SHMultiParticleH}
\mathcal{H} = \sum_{j = 0}^{N-1} h_{j j + 1} + \widetilde{h}_{N \infty}
\, .
\end{align}
The system described by this Hamiltonian is integrable. 

\section{Soft-hard open spin chain}
\label{ModelSection}

The Hamiltonian \re{SHMultiParticleH} defines a non-periodic one-dimensional lattice model of interacting noncompact spins $\bit{S}_j = (S^0_j, S^+_j, S^-_j)$ living on a semi-infinite 
light ray, with the (hard)soft boundary interaction terms determined by the SL(2) (non)invariant Hamiltonian ($\widetilde{h}_{N\infty}$)$h_{01}$. The spins form an infinite-dimensional 
representation of the sl$(2,\mathbb{R})$ algebra,
\begin{align}
[S_n^+, S_n^-] = 2 S^0_n \, , \qquad [S_n^0, S_n^\pm] = \pm S_n^\pm
\, ,
\end{align}
with an explicit representation for the action on fields at positions $z_j$ being
\begin{align}
\label{SpinGenerators}
S^+_j = z_j^2 \partial_j + 2 s z_j
\, , \qquad
S^-_j = - \partial_j
\, , \qquad
S^0_j = z_j \partial_j + s
\, ,
\end{align}
where the conformal spin $s$ labels the sl$(2,\mathbb{R})$ representations $ \mathbb{V}_j$ of a discrete series. It will be chosen to be the same for any site $j$ as well as for the soft 
boundary. The latter condition defines a homogeneous open spin chain. Its generalization to inhomogeneous case will be touched upon in the concluding section.

In our discussion, we will heavily rely on properties of functions of the light-cone coordinates analytically continued to the upper half of the complex $z$ plane with the light ray
being its boundary. Therefore, we have to introduce a proper scalar product on this space that is tailored to our needs. The inner product on the Hilbert space $\otimes^N_{j = 0} \mathbb{V}_j$ 
of $(N+1)$-variable functions holomorphic in the upper half-plane is defined as follows \cite{GelGraVil66}
\begin{align}
\label{NScalarProduct}
\VEV{\Phi|\Psi}=\int \prod_{j=0}^N D z_j\, (\Phi(z_0,\ldots,z_N))^*\, \Psi(z_0,\ldots,z_N)\,.
\end{align}
where $z_j = x_j + i y_j$ and the integration measure reads
\begin{align}
\label{MeasureSL2}
D z_j = \frac{2s - 1}{\pi}\, {dx_j d {y}_j}\, ({2y_j)^{2 s-2}} \,\theta (y_j)
\, .
\end{align}
The integration runs over the upper half-plane due to the presence of a step-function $\theta (y_j)$. The sl$(2,\mathbb{R})$ generators are antihermitian with respect to it,
\begin{align}
\label{Sconjugation}
\left( S_j^{0, \pm} \right)^\dagger = - S_j^{0, \pm}
\, , 
\end{align}
so that the Hamiltonian \re{SHMultiParticleH} is explicitly self-adjoint $\mathcal{H}^\dagger = \mathcal{H}$ yielding an orthogonal set of eigenstates.

In fact, we find it more economical to solve a unitary equivalent system obtained from the above Hamiltonian \re{SHMultiParticleH} by inversion $\mathcal{J}$. This operation
is defined at each spin-chain site $z_j$ as
\begin{align}
\label{InversionOperator}
[\mathcal{J} \Psi] (z_j) = z_j^{-2 s} \Psi (- z^{-1}_j)
\, ,
\end{align}
which, being one of the SL(2) transorfmations, leaves the inner product \re{NScalarProduct} invariant, but intertwines the sl$(2,\mathbb{R})$ generators
\begin{align}
\mathcal{J} S^{0,\pm}_j = - S^{0,\mp}_j \mathcal{J}
\, .
\end{align}
Consequently, in the inverted Hamilltonian $\mathcal{J} \mathcal{H} \mathcal{J}^{-1}$ the hard boundary is moved close to the origin $z^{-1}_\infty \to 0$, while the soft one moved to 
a large distance away. As a consequence, we find it convenient to relabel the sites in the increasing manner from the origin, i.e., $\sigma(\infty, N, N-1, \dots, 2, 1, 0) = (0, 1, 2, \dots, N, 
\infty)$,
\begin{align}
\label{InverseTotalHamiltonian}
\mathcal{H}_{\mathcal{J}} 
\equiv
\sigma \left( \mathcal{J} \mathcal{H} \mathcal{J}^{-1} \right)
=
\widetilde{h}_{01} 
+
\sum_{j = 1}^{N} h_{j j + 1}
\, ,
\end{align}
where, e.g., 
\begin{align}
\widetilde{h}_{01} = \sigma \left( \mathcal{J}\widetilde{h}_{N\infty} \mathcal{J}^{-1} \right) = \psi (1) - \ln \left( - \mu S^-_1 \right)
\, ,
\end{align}
and the soft boundary being the $(N+1)$-st site of the chain. The dynamical system determined by $\mathcal{H}_{\mathcal{J}}$ will be solved below. To get back the original one, all one has to 
do is to invert all distances in final expressions and reenumerate the sites backwards.

\section{Factorized $R$ matrices and Hamiltonians}
\label{RmatrixSection}

Our construction of a commutative system of conserved charges will be based on the existence of the Baxter operator \cite{Baxter:1971cs} and Separation of Variables (SoV) 
\cite{Skl85}. The former, in turn, will be built from intertwining operators emerging in the factorization \cite{Der05} of SL(2) invariant $R$ matrices which are the foundation 
of the Algebraic Bethe Ansatz approach to integrable systems \cite{Faddeev:1979gh}. So we will give a lightning outline of the most invaluable ingredients first.

The Lax operator, that acts on the direct product $\mathbb{C}^2 \otimes \mathbb{V}_j$ of the Hilbert space at $j$-th site $\mathbb{V}_j$ and an auxiliary two-dimensional one
$\mathbb{C}^2$, depends on the complex spectral parameter $u$ (as well as the label $s$ of the representation $\mathbb{V}_j$) and reads
\begin{align}
\label{Lax}
\mathbb{L}_j (u, s)
=
\left(
\begin{array}{cc}
u + i S^0_j & i S^-_j \\
i S^+_j        & u - i S_j^0
\end{array}
\right)
\, .
\end{align}
The product of $N+1$ copies of this operator in the auxiliary space determines the closed chain monodromy matrix $\mathbb{T} (u)$,
\begin{align}
\label{ClosedMonodromy}
\mathbb{T}_{\rm cl} (u)
=
\mathbb{L}_1 (u, s) \dots \mathbb{L}_{N+1} (u, s)
=
\left(
\begin{array}{cc}
a (u) & b (u) \\
c (u) & d (u)
\end{array}
\right)
\, ,
\end{align}
with its elements acting on the quantum space of the chain $\otimes_{j = 1}^{N+1} \mathbb{V}_j$. The open spin chain monodromy matrix is determined by ``doubling and folding''
the closed chain through the soft boundary, such that \cite{Sklyanin:1988yz}
\begin{align}
\label{OpenMonodromy}
\mathbb{T}_{\rm op} (u)
&
=
\mathbb{T}_{\rm cl} (- u)
\sigma_2 \mathbb{T}_{\rm cl}^{\rm t} (u) \sigma_2
\nonumber\\
&
=
\mathbb{L}_1 (- u, s) \dots \mathbb{L}_{N+1} (- u, s)
\mathbb{L}_{N+1} (u, s) \dots \mathbb{L}_1 (u, s)
=
\left(
\begin{array}{cc}
A (u) & B (u) \\
C (u) & D (u)
\end{array}
\right)
\, .
\end{align}
A fundamental reflection Yang-Baxter relation involving an $R$ matrix acting on the product of auxiliary spaces $\mathbb{C}^2 \otimes \mathbb{C}^2$ immediately implies that 
$B(u)$ and $C(u)$ entries form a commutative family of conserved charges,
\begin{align}
[B(u), B(v)] = [C(u), C(v)] = 0
\, ,
\end{align}
while $A$ and $D$ are not individually, but only in the sum. Since the $B$-entry of the open spin chain monodromy matrix will play a distinguished role in our consideration below,
let us a point out a few of its salient properties. Making use of the first definition in Eq.\ \re{OpenMonodromy}, one finds its relation to the elements of the closed chain monodromy
\begin{align}
\label{Brel}
B (u) = b (- u) a(u) - a(-u) b(u)
\, .
\end{align}
From Eq.\ \re{Brel} and the conjugation property \re{Sconjugation}, it is straightforward to verify that
\begin{align}
\label{Bconjugation}
\big( B (u) \big)^\dagger = - B (- u^\ast)
\end{align}
as a consequence of $\big( a (u) \big)^\dagger = a(u^\ast)$ and $\big( b (u) \big)^\dagger = b (u^\ast)$.  Finally, from the definition \re{OpenMonodromy}, it follows that $B (u)$ is an 
operator polynomial in $u$ of degree $2N + 1$. However, it possesses a kinematic zero at $u = - i/2$ as was shown in Ref.\ \cite{Derkachov:2003qb} and found explicitly by different 
means in Section \ref{SectionBaxterEq} below. Then the operator can be decomposed as
\begin{align}
\label{BopenOperatorZeros}
B (u) = (-1)^N (2u + i) i S^- \prod_{j = 1}^{N} (u^2 - \widehat{x}_j^2) 
\, ,
\end{align}
in terms of $N$ operator zeros $ \widehat{x}_j$, i.e., the Separated Variables.

The pairwise Hamiltonians defining the open chain arise, on the other hand, from the SL(2) invariant  $R$ matrices acting on the product of noncompact quantum spaces and obey an 
$RLL$ relation
\begin{align}
\label{Rcheck}
\check{\mathcal{R}}_{jk} (u - v)\, \mathbb{L}_j (u,s_j) \mathbb{L}_k (v,s_k)
=
\mathbb{L}_j (v,s_j) \mathbb{L}_k (u,s_k)\, \check{\mathcal{R}}_{jk} (u-v)
\, ,
\end{align}
where one conventionally pulls out a permutation operator $\Pi_{jk}$ acting on the product of two spaces, ${\mathcal{R}}_{jk} =\Pi_{jk} \check{\mathcal{R}}_{jk}$. It is this operator
that was found to factorize in terms of intertwiners $\mathcal{R}^\pm$ \cite{Der05} 
\begin{align}
\check{\mathcal{R}}_{jk} (u) = \mathcal{R}^{+}_{jk} (\gamma_{kj}) \mathcal{R}^{-}_{jk} (\gamma_{jk})
\, ,
\end{align}
which depend on a linar combination of the spectral parameter and spins $\gamma_{jk} = s_j - s_k + i u$, 
\begin{align}
\label{RminusTwoSites}
\mathcal{R}^-_{jk} (\gamma) 
= 
\mathcal{R}^+_{kj} (\gamma) 
= 
\frac{\Gamma (2s_j)}{\Gamma (2 s_j - \gamma)} 
\frac{\Gamma (z_{jk} \partial_j + 2 s_j - \gamma)}{\Gamma (z_{jk} \partial_j + 2 s_j )}
\, .
\end{align}
These operators intertwine the quantum spaces of the chain in the following fashion\footnote{Here for clarity, we temporarily introduced different conformal spins $s_j$ for all sites and used them 
to label quantum spaces.}
\begin{align}
\mathcal{R}^\mp_{jk} (\gamma): \mathbb{V}_{s_j} \otimes  \mathbb{V}_{s_k} \to \mathbb{V}_{s_j \mp \gamma/2} \otimes  \mathbb{V}_{s_k \pm \gamma/2} 
\, ,
\end{align}
such that the original ${\mathcal{R}}_{jk}$ maps $\mathbb{V}_{s_j} \otimes  \mathbb{V}_{s_k} \to \mathbb{V}_{s_j} \otimes  \mathbb{V}_{s_k}$.

As can be easily verified, the expansion of $\mathcal{R}^\mp$ in the vicinity of $\gamma = 0$ generates the bulk pairwise Hamiltonians (including the one for the interaction with the 
soft boundary),
\begin{align}
\mathcal{R}^\mp_{jk} (\gamma) = 1 + \gamma \left( h_{jk}^\mp + \psi (2s) - \psi(1) \right) + O (\gamma^2)
\, ,
\end{align}
where
\begin{align}
h_{jk}^- = \psi(2s) - \psi (z_{jk} \partial_j + 2s)
\, , \qquad
h_{jk}^+ = \psi(2s) - \psi (z_{kj} \partial_k + 2s)
\, ,
\end{align}
such that $h_{jk} =  h_{jk}^-  +  h_{jk}^+ + 2 \psi (1) - 2 \psi (2s)$ with $h_{jk}$ introduced in Eq.\ \re{PairwiseHamiltonians}. While the Hamiltonian for the interaction with the hard boundary 
emerges from a limit of the bulk $R$ matrix. Namely, taking $z_k \to \infty$, we find
\begin{align}
\label{RminusOneSite}
\mathcal{R}^-_j (\gamma) \equiv \lim_{z_k \to \infty} {\rm e}^{i \pi \gamma} z_j^{2 s} \mathcal{R}_{jk} (\gamma) 
= 
\frac{\Gamma (2s)}{\Gamma (2s - \gamma)} \partial_j^{- 2 \gamma}
\, ,
\end{align}
with the small-$\gamma$ expansion producing
\begin{align}
\mathcal{R}^-_j (\gamma)
=
1 + \gamma \big(\, \widetilde{h}_{0j} + \psi (2s) - \psi(1) \big) + O (\gamma^2)
\, .
\end{align}
Possessing this knowledge, let us move on to the construction of the Baxter operator and prove its commutativity with certain elements of the monodromy matrix \re{OpenMonodromy}.

\section{Baxter operator}
\label{BaxterQsection}

Within the context of the Hamiltonian system \re{InverseTotalHamiltonian}, the Baxter operator $\mathbb{Q}$ maps the open spin chain into itself  $\otimes_{j = 1}^{N+1} \mathbb{V}_j 
\to \otimes_{j = 1}^{N+1} \mathbb{V}_j$ and obeys the properties:
\begin{itemize}
\item Baxter equation
\begin{align}
\label{B-BaxterEq}
B (u) \mathbb{Q} (u) = (-1)^{N} (2u + i) (u + i s)^{2N + 1}  \mathbb{Q} (u + i)
\, , 
\end{align}
\item Commutativity conditions
\begin{align}
\label{Qcommutator}
[\mathbb{Q} (u), \mathbb{Q} (v)] = 0
\end{align}
and
\begin{align}
\label{BQcommutator}
[B (u), \mathbb{Q} (v)] = 0
\, .
\end{align}
\end{itemize}
Its construction can be systematically accomplished making use of intertwining relations for the $\mathcal{R}^\pm$ operators as was done, for instance, in Ref.\ \cite{Belitsky:2014rba}
for the hard-hard open spin chains. However, we will not follow this route in the current presentation and rely instead on a diagrammatic technique introduced in Ref.\ \cite{Derkachov:2002tf}.

\begin{figure}[t]
\begin{center}
\mbox{
\begin{picture}(0,100)(190,0)
\put(0,-440){\insertfig{27}{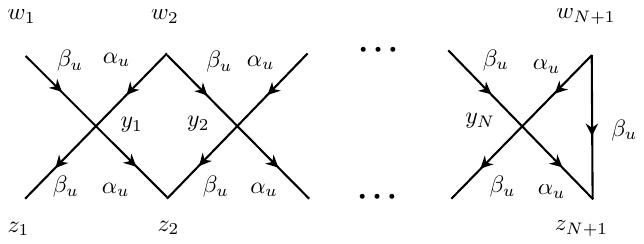}}
\end{picture}
}
\end{center}
\caption{ \label{BBaxterFig} Representation for the kernel of the Baxter kernel.
}
\end{figure}

Motivated by our findings at the end of the last section, let us consider the following `doubled and folded' chain of $\mathcal{R}^-$ operators of the argument $\gamma = \alpha_u \equiv s + i u$
\begin{align}
\label{BBaxterOperator}
\mathbb{Q} (u) 
= 
\mathcal{R}^-_{12} (\alpha_u) \mathcal{R}^-_{23} (\alpha_u) \dots \mathcal{R}^-_{NN+1} (\alpha_u)  
\mathcal{R}^-_{N+1N} (\alpha_u) \dots  \mathcal{R}^-_{32} (\alpha_u) \mathcal{R}^-_{21} (\alpha_u) 
\mathcal{R}^-_{1} (\alpha_u)
\, ,
\end{align}
with $\mathcal{R}^-_{jk}$ and $\mathcal{R}^-_j$ defined in Eqs.\ \re{RminusTwoSites} and \re{RminusOneSite}, respectively. To start with, let us find an integral kernel corresponding to it. 
The latter can be put in correspondence to any operator $\mathbb{A}$ acting on the Hilbert space of the chain and can be associated to a function $\mathcal{A}$ of $N+1$ holomorphic and 
$N+1$ anti-holomorphic variables in a unique way via the relation
\begin{align}
\label{integral-kernel}
[\mathbb{A} \Psi] (z_0,\ldots,z_N)=\int\prod_{k=1}^{N+1} D w_k\, \mathcal{A} (z_1,\ldots, z_{N + 1}|\bar w_1,\ldots,\bar w_{N + 1}) \Psi(w_1,\ldots,w_{N+1})\,.
\end{align}
A straightforward calculation making use of the integral representation for the Euler Beta function and basic integrals from, e.g., Appendix A of Ref.\ \cite{Belitsky:2014rba}, allows us
to cast the kernel $\mathcal{Q}_u$ of the $\mathbb{Q} (u)$ into the form
\begin{align}
\mathcal{Q}_u (z_1, \dots, & z_{N+1}
| \bar{w}_1 \dots , \bar{w}_{N+1})
=
{\rm e}^{i \pi s (2N + 1)}
\int \prod_{j = 1}^{N} D_s y_j \, (z_1 - \bar{y}_1)^{- \beta_u}
\\
&\times
Y_u (z_2, \dots, z_{N}, z_{N+1} | \bar{y}_1, \dots, \bar{y}_{N}, \bar{w}_{N+1})
Y_{-u} (y_1, \dots, y_{N-1}, y_{N} | \bar{w}_1, \dots, \bar{w}_{N}, \bar{w}_{N+1})
\, ,
\nonumber
\end{align}
where we introduced the function \cite{Derkachov:1999pz}
\begin{align}
\label{YyFunctions}
Y_u (z_1, \dots, z_{N-1}, z_{N} | \bar{w}_1, \dots, \bar{w}_{N}, \bar{w}_{N+1})
=
\prod_{j=1}^{N} y_u (z_j | \bar{w}_j , \bar{w}_{j+1}) 
\, , 
\end{align}
with individual factors in it being
\begin{align}
y_u (z_j | \bar{w}_j , \bar{w}_{j+1}) =  (z_j - \bar{w}_j)^{- \alpha_u} (z_j - \bar{w}_{j+1})^{- \beta_u}
\, ,
\end{align}
and $\alpha_u \equiv s + i u$ and $\beta_u = s - i u$. The kernel of the Baxter operator is shown in Fig. \ref{BBaxterFig} as a two-dimensional Feynman graph with the 
propagator from $w$ to $z$ defined by
\begin{figure}[h]
\begin{center}
\mbox{
\begin{picture}(0,10)(90,0)
\put(0,-197){\insertfig{10}{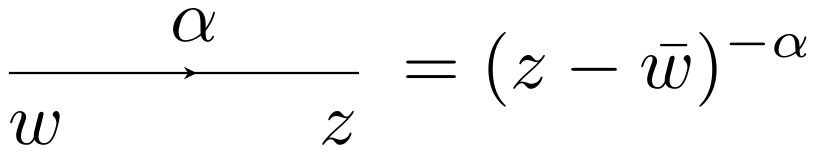}}
\end{picture}
}
\end{center}
\end{figure}

\begin{figure}[t]
\begin{center}
\mbox{
\begin{picture}(0,345)(250,0)
\put(0,-40){\insertfig{18}{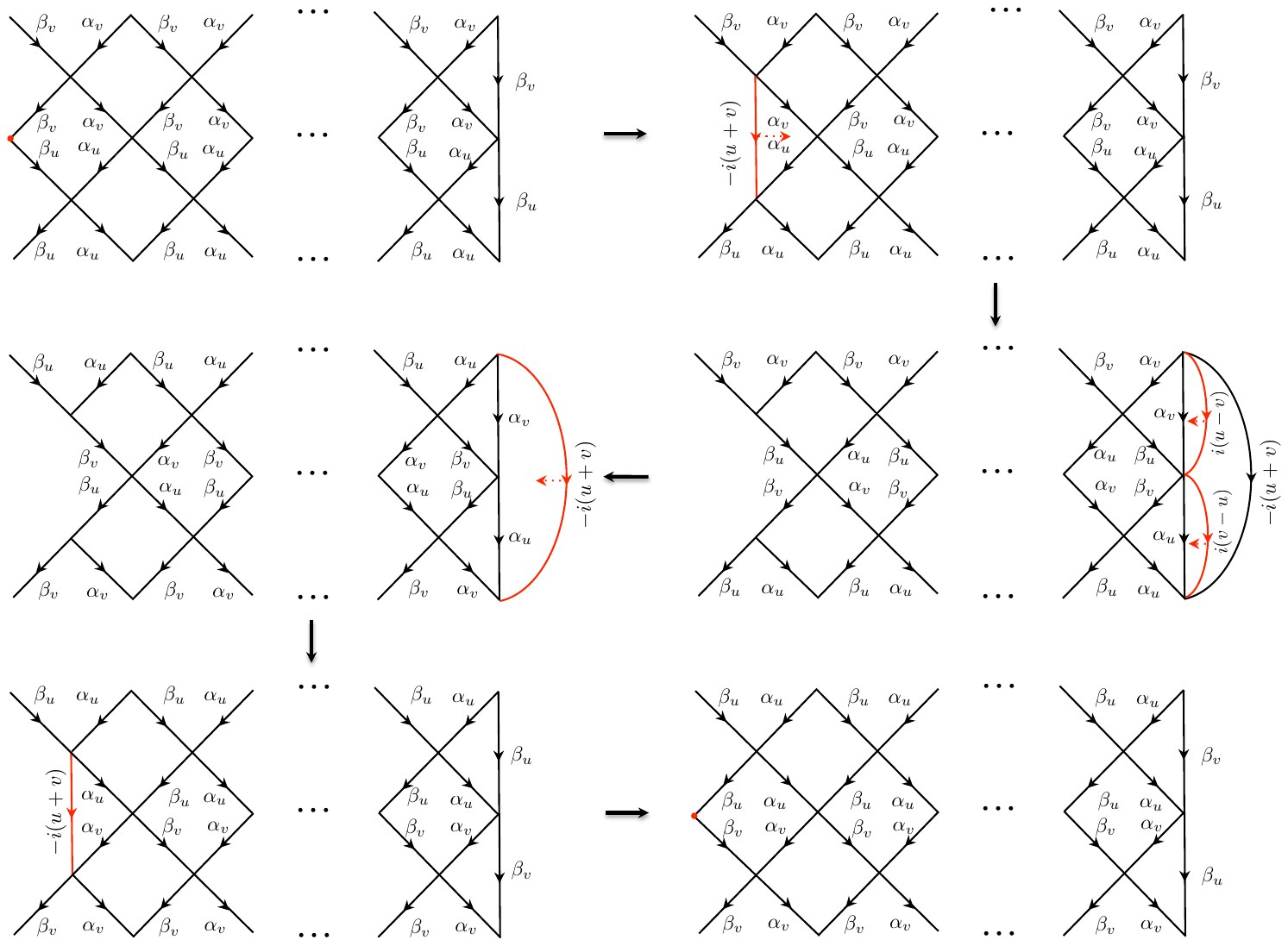}}
\end{picture}
}
\end{center}
\caption{ \label{BaxterCommFig} Proof of the commutativity of the Baxter operators.
}
\end{figure}

The commutativity of the Baxter operators for different values of the spectral parameter follows immediately from the diagrammatic representation of the their product
and is shown by the moves in the sequence of graphs in Fig.\ \ref{BaxterCommFig}. Namely, first, one integrates out the the leftmost vertex (see the top left graph in Fig.\ 
\ref{BaxterCommFig}) connecting the two Baxter kernels via the chain rule given in Appendix \ref{FeynmanRulesAppendix}. Then one moves (see the top right graph)
the vertical propagator from left to right via the permutation identity from Appendix \ref{FeynmanRulesAppendix}. At the next step, one splits the labels of the two rightmost 
lines within each Baxter kernel as $\alpha_{u/v} = \alpha_{v/u} \pm i (u - v)$ and moves the $\pm i (u - v)$-propagators all the way to the left with the same permutation
identity (as shown in the middle right panel). After that, one shifts the remaining propagator, left over from step one, to the left as well (left middle graph). As a result, one
ends up with the left diagram in the bottom row of Fig.\ \ref{BaxterCommFig}. Finally, reconstructing two propagators from one by using the chain rule backwards, we get
the right bottom graph, where compared to the one we started from, the $u$ and $v$ parameters are interchanged. This completes the verification of Eq.\ \re{Qcommutator}.

The commutativity of $\mathbb{Q}$ and $B$ immediately follows from the Baxter relation \re{B-BaxterEq} by taking the hermitian conjugate of both of its sides. Namely,
the left-hand side gives $- \mathbb{Q} (- u^\ast) B(- u^\ast)$, where we used the property \re{Bconjugation}. A conjugate of the right-hand side of the Baxter relation 
yields $(2u^\ast - i) (u^\ast - i s)^{2N+1} Q (u^\ast - i) = - B(- u^\ast) \mathbb{Q} (- u^\ast)$. Equating the results, immediately confirms Eq.\ \re{BQcommutator}. Thus everything 
boils down to establishing Eq.\ \re{B-BaxterEq}. We will turn to its proof next. 

Before we come to this, we close this section with a relation of the open spin chain Hamiltonian to the Baxter operator. Namely, the former is determined by the logarithmic 
derivative of $\mathbb{Q} (u)$ at $u = i s$,
\begin{align}
\mathcal{H}_{\mathcal{J}} = - i \big( \ln \mathbb{Q} (i s) \big)^\prime + (2 N + 1) \big( \psi (1) - \psi (2s) \big)
\, ,
\end{align}
with $\mathcal{H}_{\mathcal{J}}$ of Eq.\ \re{InverseTotalHamiltonian}.

\section{Baxter equation}
\label{SectionBaxterEq}

To establish the Baxter equation \re{B-BaxterEq} for the operator \re{BBaxterOperator}, we will use the Gaudin-Pasquier trick \cite{PasGau92}. It relies on transformation 
properties of the elements of the monodromy matrix under a gauge transformation of the Lax operators,
\begin{align}
\mathbb{L}_j (u,s) \to \mathbb{L}^\prime_j (u, s; \bar{w}_j, \bar{w}_{j+1}) = \mathbb{M}_j^{-1} \mathbb{L}_j (u,s) \mathbb{M}_{j+1}
\, .
\end{align}
It will be convenient to choose $\mathbb{M}_j$ in the form
\begin{align}
\mathbb{M}_j
=
\left(
\begin{array}{cc}
1 & \bar{w}_j^{-1} \\
0 & 1
\end{array}
\right)
\, ,
\end{align}
such that it goes to the identity matrix as the gauge parameter is sent to infinity, $\bar{w}_j \to \infty$. The calculation of the elements of $ \mathbb{L}^\prime_j$ is simplified making use of 
the lower-triangular factorization of the Lax operator,
\begin{align}
\mathbb{L}_j (u, s)
=
i z_j^{- \alpha_u - \beta_u}  
\left(
\begin{array}{cc}
1  & z_j^{-1} \\
0  & 1
\end{array}
\right)
\left(
\begin{array}{cc}
- \alpha_u         & 0 \\
z_j^2 \partial_j  & - 1 + \beta_u
\end{array}
\right)
\left(
\begin{array}{cc}
1  & - z_j^{-1} \\
0  & 1
\end{array}
\right)
 z_j^{\alpha_u + \beta_u}
\, .
\end{align}
Instead of listing explicit elements, let us demonstrate their action on the function $Y_u$ introduced in the previous section. In fact, it was introduced
as a function that is annihilated by $[\mathbb{L}^\prime_j (u, s)]_{12}$, such that
\begin{align}
\label{Gaugea}
[\mathbb{L}^\prime_j (u, s; \bar{w}_j, \bar{w}_{j+1})]_{11} \, 
y_u (z_j | \bar{w}_j, \bar{w}_{j+1})
&
= (u+is) \frac{\bar{w}_{j+1}}{\bar{w}_j} y_{u + i}(z_j | \bar{w}_j, \bar{w}_{j+1})
\, , \\
\label{Gaugeb}
[\mathbb{L}^\prime_j (u, s; \bar{w}_j, \bar{w}_{j+1})]_{12} \, 
y_u (z_j | \bar{w}_j, \bar{w}_{j+1})
&
=
0
\, , \\
[\mathbb{L}^\prime_j (u, s; \bar{w}_j, \bar{w}_{j+1})]_{21} \,  
y_u (z_j | \bar{w}_j, \bar{w}_{j+1})
&
= - \frac{\partial}{\partial z_j^{-1}} y_{u}(z_j | \bar{w}_j, \bar{w}_{j+1})
\, , \\
[\mathbb{L}^\prime_j (u, s; \bar{w}_j, \bar{w}_{j+1})]_{22} \, 
y_u (z_j | \bar{w}_j, \bar{w}_{j+1})
&
= (u-is) \frac{\bar{w}_j}{\bar{w}_{j + 1}} y_{u - i}(z_j | \bar{w}_j, \bar{w}_{j+1})
\, .
\end{align}

Relying on the first line in Eq.\ \re{OpenMonodromy}, we find that the elements of the open spin chain monodromy matrix depend only on the gauge parameter $\bar{w}_0$.
Since we focus, for obvious reasons, on the $B$-element, let us introduce a two spectral-parameter function
in particular,
\begin{align}
\label{GaugeTransB}
B^\prime (u, v; w_0) = b^\prime (v; \bar{w}_0) a^\prime (u; \bar{w}_0) - a^\prime (v; \bar{w}_0)  b^\prime (u; \bar{w}_0) 
\, ,
\end{align}
such that
\begin{align}
B (u) 
=
\lim_{\bar{w}_0 \to \infty} B^\prime (u, -u; w_0) 
\, .
\end{align}
This is a crucial property which we will explore in our subsequent derivation.

\begin{figure}[t]
\begin{center}
\mbox{
\begin{picture}(0,240)(200,0)
\put(0,-300){\insertfig{27}{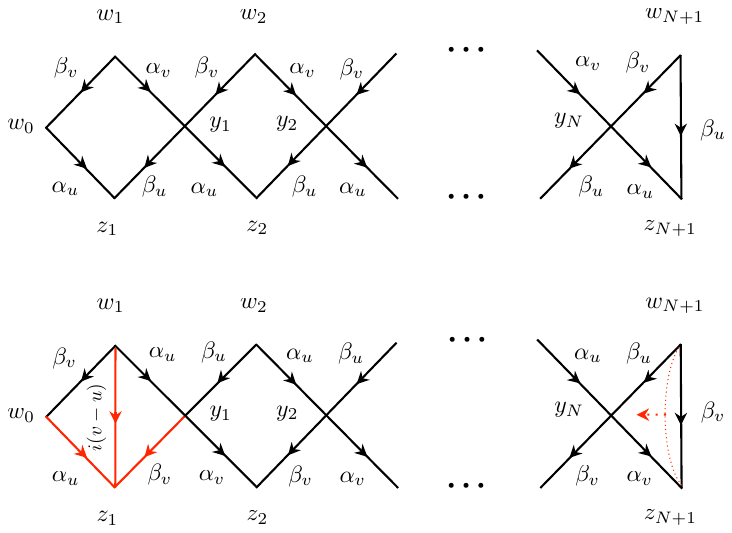}}
\end{picture}
}
\end{center}
\caption{ \label{WFig} Graphical representation for the auxiliary function $W_{u,v}$ (top) and its transformed form (bottom) after splitting the
rightmost vertical line as $\beta_u = \beta_v + i (v-u)$ and  moving the line with the index $i(v-u)$ all the way to the left till it lands in the red
subgraph.
}
\end{figure}

Now, we introduce an auxiliary function
\begin{align}
\mathcal{W}_{u,v} (z_1, &
\dots, z_{N+1} | \bar{w}_0, \dots, \bar{w}_{N+1})
=
{\rm e}^{i \pi s (2N + 1)}
\int \prod_{j = 1}^{N} D_s y_j \, (w_0 - \bar{w}_1)^{- \beta_v}
\\
&\times
Y_u (z_1, \dots, z_{N}, z_{N+1} | \bar{w}_0, \bar{y}_1, \dots, \bar{y}_{N}, \bar{w}_{N+1})
Y_{v} (y_1, \dots, y_{N-1}, y_{N} | \bar{w}_1, \dots, \bar{w}_{N}, \bar{w}_{N+1})
\, , \nonumber
\end{align}
with its diagrammatic realization shown in Fig.\ \ref{WFig}. One can immediately see, as a result of Eq.\ \re{Gaugeb}, that
\begin{align}
b^\prime (u; \bar{w}_0) \mathcal{W}_{u,v} (z_1, \dots, z_{N+1} | \bar{w}_0, \dots, \bar{w}_{N+1}) = 0
\, ,
\end{align}
so that the second term in the definition of $B^\prime (u, v; w_0)$ in Eq.\ \re{GaugeTransB} does not contribute, and we find in this manner
\begin{align}
B^\prime (u, v; w_0) 
\mathcal{W}_{u,v} (z_1, \dots, z_{N+1} | \bar{w}_0, \dots, \bar{w}_{N+1})
&
=
(u+ is)^{N+1} \frac{\bar{w}_{N+1}}{\bar{w}_0} 
\\
&\times
b^\prime (v; \bar{w}_0) \mathcal{W}_{u + i,v} (z_1, \dots, z_{N+1} | \bar{w}_0, \dots, \bar{w}_{N+1})
\, , \nonumber
\end{align}
where Eq.\ \re{Gaugea} was applied.

To calculate the result of the action of $b^\prime (v; \bar{w}_0)$ in the most efficient manner, let us use the permutation identity by moving the propagator
$(z_{N+1} - \bar{w}_{N+1})^{i (u - v)}$ from right to left, such that the result for $\mathcal{W}_{u + i,v} (z_1, \dots, z_{N+1} | \bar{w}_0, \dots, \bar{w}_{N+1})$ now
reads
\begin{align}
\label{WafterTheMove}
\mathcal{W}_{u  + i,v} (z_1, &
\dots, z_{N+1} | \bar{w}_0, \dots, \bar{w}_{N+1})
=
{\rm e}^{i \pi s (2N + 1)}
\int \prod_{j = 1}^{N} D_s y_j \, (w_0 - \bar{w}_1)^{- \beta_v} \widetilde{y}_{u + i,v} (z_1|\bar{w_0}, \bar{w}_1, \bar{y}_1)
\\
&\times
Y_v (z_2, \dots, z_{N}, z_{N+1} | \bar{y}_1, \dots, \bar{y}_{N}, \bar{w}_{N+1})
Y_{u+i} (y_1, \dots, y_{N-1}, y_{N} | \bar{w}_1, \dots, \bar{w}_{N}, \bar{w}_{N+1})
\, , \nonumber
\end{align}
and it is shown explicitly in Fig.\ \ref{WFig}, with a combination of the propagators $\widetilde{y}_{u,v} (z_1|\bar{w_0}, \bar{w}_1, \bar{y}_1)$ designated by the red subgraph,
\begin{align}
\widetilde{y}_{u,v} (z_1|\bar{w_0}, \bar{w}_1, \bar{y}_1)
\equiv
(z_1 - \bar{w}_0)^{- \alpha_u} (z_1 - \bar{w}_1)^{i (u - v)} (z_1 - \bar{y}_1)^{- \beta_v}
\, .
\end{align}
The action of $b^\prime (v; \bar{w}_0)$ on the integrand $\mathcal{W}_{u  + i,v} $, again thanks to Eq.\ \re{Gaugeb}, factorizes as
\begin{align}
b^\prime (v; \bar{w}_0)
\widetilde{y}_{u + i,v}  Y_v Y_{u+i}
=
(v - i s)^{N} \frac{\bar{y}_1}{\bar{w}_{N+1}}  Y_{v - 1} Y_{u+i} 
[\mathbb{L}^\prime_j (u, s; \bar{w}_0, \bar{y}_1)]_{12} \, \widetilde{y}_{u + i,v} 
\, , 
\end{align}
where, for brevity, we did not display the arguments of the functions involved, but they can easily be read off from Eq.\ \re{WafterTheMove}. Finally,
\begin{align}
[\mathbb{L}^\prime_j (u, s; \bar{w}_0, \bar{y}_1)]_{12} \, \widetilde{y}_{u + i,v} (z_1|\bar{w_0}, \bar{w}_1, \bar{y}_1)
=
(v - u - i) \frac{\bar{w}_0 - \bar{w}_1}{\bar{w}_0 \bar{y}_1}
\widetilde{y}_{u + i,v - i} (z_1|\bar{w_0}, \bar{w}_1, \bar{y}_1)
\, .
\end{align}
Combining all results together, we find that the auxiliary function obeys the following equation
\begin{align}
\label{Wequation}
B^\prime (u, v; w_0) \mathcal{W}_{u,v} (z_1, \dots, z_{N+1} | \bar{w}_0, \dots, \bar{w}_{N+1})
&=
(u+is)^{N+1} (v - i s)^{N} (v - u - i)
\\
&\times\frac{\bar{w}_0 - \bar{w}_1}{\bar{w}_0^2 (w_0 - \bar{w}_1)}
\mathcal{W}_{u + i,v - i} (z_1, \dots, z_{N+1} | \bar{w}_0, \dots, \bar{w}_{N+1})
\, . \nonumber
\end{align}
Taking the limit $w_0 \to \infty$ with a proper scaling factor, we uncover the kernel of the Baxter operator
\begin{align}
\mathcal{Q}_u (z_1, \dots, z_{N+1} | \bar{w}_0, \dots, \bar{w}_{N+1})
=
\lim_{|w_0| \to \infty} ( - w_0 \bar{w}_0)^{\alpha_u} \mathcal{W}_{u,-u} (z_1, \dots, z_{N+1} | \bar{w}_0, \dots, \bar{w}_{N+1})
\, ,
\end{align}
and the Baxter equation itself \re{B-BaxterEq}.

\begin{figure}[t]
\begin{center}
\mbox{
\begin{picture}(0,100)(190,0)
\put(0,-440){\insertfig{27}{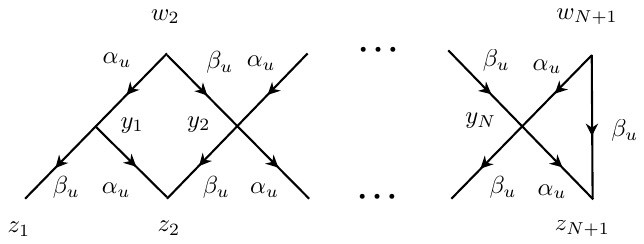}}
\end{picture}
}
\end{center}
\caption{ \label{LayerFig} Graphical representation for the layer kernel $\Lambda_u$.
}
\end{figure}

Since the Baxter equation is a one-term recursion relation, it can be solved in a straightforward fashion. However, an overall normalization constant and
a periodic function $f(u+i) = f(u)$ remain arbitrary. We will fix both of them in the next section by explicitly computing the eigenvalues of $\mathbb{Q} (u)$.
The result of the analysis which follows is summarized in the equation
\begin{align}
\label{SolutionBaxter}
\mathbb{Q} (u) = 
\left( (S^-)^{- i u} \prod_{j=1}^N \Gamma (- i u - i \widehat{x}_j) \Gamma (- i u + i \widehat{x}_j) \right) / \Gamma^{2N + 1} (- i u + s)
\, ,
\end{align}
where we used the representation of $B (u)$ in terms of its operator zeros \re{BopenOperatorZeros}.

\section{Eigenfunctions}
\label{EigenfunctionsSection}

As it is clear from Eq.\ \re{Wequation} that, if in addition to sending $|w_0| \to \infty$, we would follow it up by $\bar{w}_1 \to \infty$, we immediately uncover that 
\begin{align}
B (u) \Lambda_u (z_1, \dots, z_{N+1} | w_2, \dots, w_{N+1}) = 0
\, ,
\end{align}
where
\begin{align}
\Lambda_u (z_1, \dots, z_{N+1} | w_2, \dots, w_{N+1})
=
\lim_{\bar{w}_1 \to \infty} \bar{w}_1^{\alpha_u} \mathcal{Q}_u (z_1, \dots, z_{N+1} | w_1, \dots, w_{N+1})
\, ,
\end{align}
with the kernel given by
\begin{align}
\Lambda_u (z_1, \dots, &z_{N+1} | w_2, \dots, w_{N+1}) 
= 
{\rm e}^{i \pi s (2N + 1)}
\int \prod_{j = 1}^{N} D_s y_j \, (z_1 - \bar{y}_1)^{- \beta_u} (y_1 - \bar{w}_2)^{- \alpha_u}
\\
&\times
Y_u (z_2, \dots, z_{N}, z_{N+1} | \bar{y}_1, \dots, \bar{y}_{N}, \bar{w}_{N+1})
Y_{- u} (y_1, \dots, y_{N-1}, y_{N} | \bar{w}_2, \dots, \bar{w}_{N}, \bar{w}_{N+1})
\, , \nonumber
\end{align}
shown in Fig.\ \ref{LayerFig}. This is nothing else as the defining equation for the so-called layer kernel of the open spin chain \cite{Derkachov:2003qb} (see also recent \cite{Braun:2018fiz}).
It is now straightforward to recursively construct the eigenfunction that diagonalizes the $B$ operator by stacking these layers up with their labels  $\bit{x} = (x_1, \dots, x_N)$ 
determined by the eigenvalues of its operators zeros,
\begin{align}
\widehat{x}_j \Psi_{p,\bit{\scriptstyle x}} (z_1, \dots, z_{N+1}) = x_j \Psi_{p,\bit{\scriptstyle x}} (z_1, \dots, z_{N+1})
\, .
\end{align}

\begin{figure}[t]
\begin{center}
\mbox{
\begin{picture}(0,170)(250,0)
\put(0,-350){\insertfig{25}{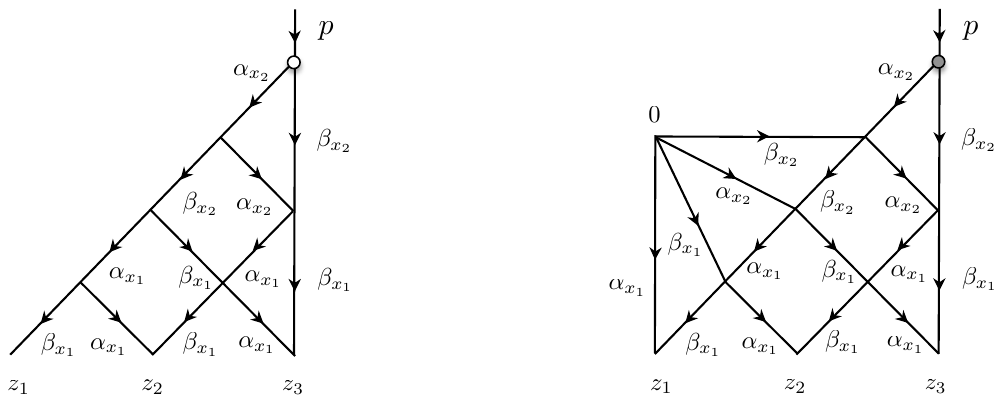}}
\end{picture}
}
\end{center}
\caption{ \label{WaveFunctionsFig} Graphical representation for the wave functions of the operators $B(u)$ (left) and $C(u)$ (right).
}
\end{figure}

\begin{figure}[t]
\begin{center}
\mbox{
\begin{picture}(0,50)(155,0)
\put(0,-390){\insertfig{20}{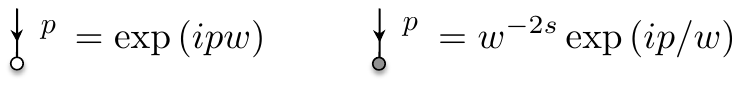}}
\end{picture}
}
\end{center}
\caption{ \label{TopVertexFig} Feynman diagrams for the vertex of the factor of the wave functions for the operators $B(u)$ (left) and $C(u)$ (right).
}
\end{figure}

Explicitly \cite{Derkachov:2003qb},
\begin{align}
\Psi_{p,\bit{\scriptstyle x}} (z_1, \dots, z_{N+1})
&
=
\int \prod_{j=2}^{N+1} D_s w^{(N)}_j \Lambda_{x_1} (z_1, \dots, z_{N+1} | w^{(N)}_2, \dots, w^{(N)}_{N+1})
\nonumber\\
&
\times
\int \prod_{j=3}^{N+1} D_s w^{(N-1)}_j \Lambda_{x_2} (w^{(N)}_2, \dots, w^{(N)}_{N+1} | w^{(N-1)}_3, \dots, w^{(N-1)}_{N+1})
\nonumber\\
&\ \,
\vdots
\nonumber\\
&
\times
\int D_s w^{(1)}_{N+1} \Lambda_{x_N} (w^{(2)}_N, w^{(2)}_{N+1} | w^{(1)}_{N+1})
\exp \left( i p \, w^{(1)}_{N+1} \right)
\nonumber\\
\end{align}
demonstrated graphically in the left panel of Fig.\ \ref{WaveFunctionsFig}, where the top is crowned by the plane wave (see the left panel in Fig.\ \ref{TopVertexFig}) with 
the eigenvalue $p$ of $S^- = - \sum_{j = 1}^{N + 1} \partial_j$,
\begin{align}
i S^- \Psi_{p,\bit{\scriptstyle x}} (z_1, \dots, z_{N+1}) = p \, \Psi_{p,\bit{\scriptstyle x}} (z_1, \dots, z_{N+1})
\, .
\end{align}

\begin{figure}[p]
\begin{center}
\mbox{
\begin{picture}(0,580)(260,0)
\put(0,200){\insertfig{20}{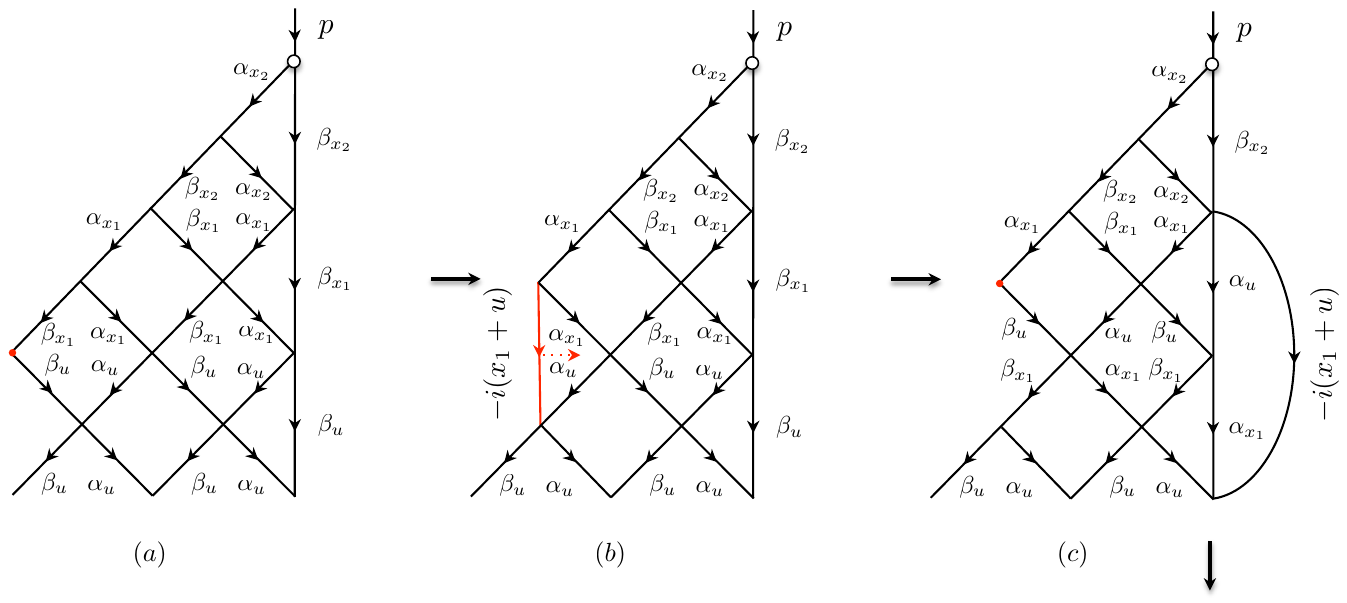}}
\put(0,-10){\insertfig{20}{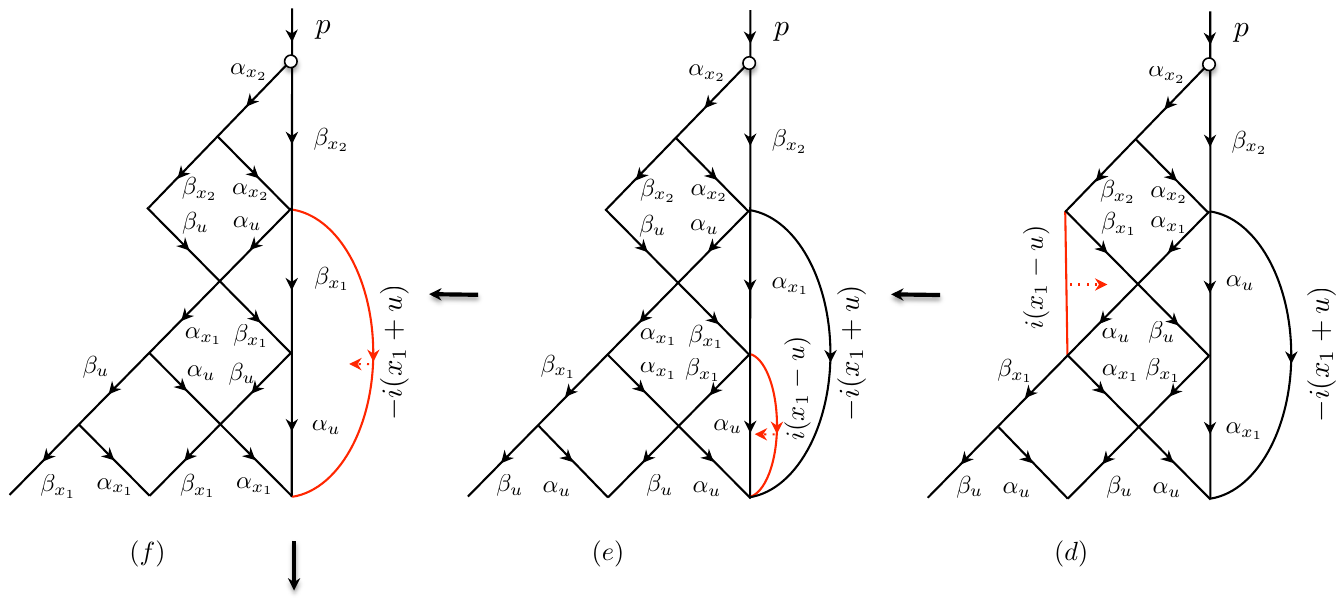}}
\put(0,-220){\insertfig{20}{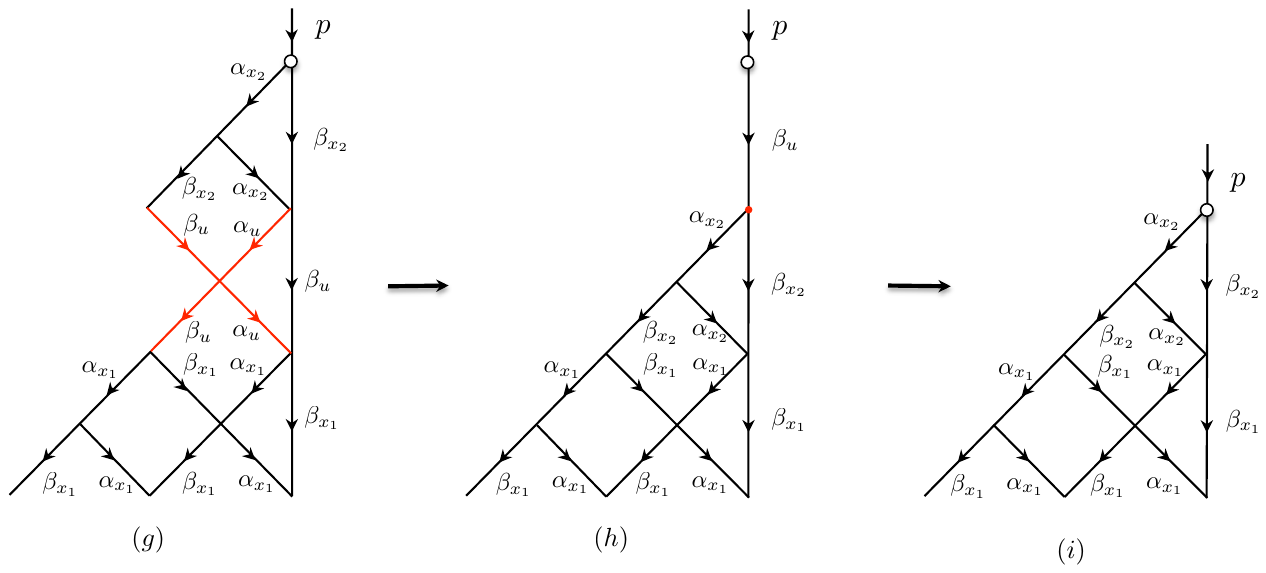}}
\end{picture}
}
\end{center}
\caption{ \label{QeigenfunctionFig} Calculation of eigenfunctions of the Baxter operator.
}
\end{figure}

With this formula, one can immediately find the eigenvalues of the Baxter operator in a recursive fashion. We exemplify it in Fig.\ \ref{QeigenfunctionFig} for $N=2$.
To start with, one integrates the leftmost vertex in the product of the Baxter kernel and the wave function (see the top left graph in Fig.\ \ref{QeigenfunctionFig}) 
making use of the chain rule giving the middle graph in the top row multiplied by
\begin{align}
{\rm e}^{- i \pi s} a (\beta_u, \beta_{x_1})
\, ,
\end{align}
with $a$ given in Eq.\ \re{aChain}. Then, one moves the vertical propagator relying on the permutation identity to the rightmost position (right top panel). Next,
one repeats the same for the remaining leftmost vertex of this layer acquiring the factor
\begin{align}
{\rm e}^{- i \pi s} a (\beta_u, \alpha_{x_1})
\end{align}
along the way, and then moving this propagator to the right as shown in the rightmost figure in the middle row of Fig.\ \ref{QeigenfunctionFig}. At a subsequent step,
the label on the rightmost vertical propagator is decomposed as $\alpha_{x_1} = \alpha_u + i (x_1 - u)$, with the propagator $i (x_1 - u)$ moved leftmost as in
the middle panel of the middle row. Finally, to complete this layer, one moves the overarching propagator $- i (x_1 + u)$, remaining from the first step, all the way to the 
left again. One ends up with the graph in the left of the bottom row of Fig.\ \ref{QeigenfunctionFig}. We see that after all of these steps, one ends up with the Baxter
kernel (shown by the red subgraph) acting on a layer of wave-function with one site less than we started from. Repeating all of the above all over again, we get the factor
\begin{align}
{\rm e}^{- 2 i \pi s} a (\beta_u, \beta_{x_2}) a (\beta_u, \alpha_{x_2})
\, ,
\end{align}
multiplying the middle bottom diagram. Computing the remaining Fourier integral with the help of Eq.\ \re{FourierTransform}, we get the rightmost graph, which is nothing 
else as the $N=2$ wave function multiplied by
\begin{align}
\frac{\Gamma (2s)}{\Gamma (\beta_u)} {\rm e}^{- i \pi \beta_u/2} p^{- \alpha_u}
\, .
\end{align}
Combining everything together in this manner, we establish Eq.\ \re{SolutionBaxter}, where the operators are replaced by their eigenvalues for $N=2$.

The proof of the orthogonality of $\Psi_{p,\bit{\scriptstyle x}}$ can, again, be accomplished recursively. However, we spare the reader the details since they can be found in 
Ref.\ \cite{Derkachov:2003qb} and merely quote the final result. The scalar product reads
\begin{align}
\vev{\Psi_{p',\bit{\scriptstyle x}'} | \Psi_{p,\bit{\scriptstyle x}}}
=
\mu (\bit{x})
\delta (p' - p) \sum_\sigma \delta^N (\bit{x}'_\sigma - \bit{x})
\, ,
\end{align}
where the sum stands for all permutations of $N$ eigenvalues and the measure reads
\begin{align}
\mu (\bit{x}) 
&= 
(2\pi)^N \Gamma^{N+1} (2s) \prod_{j=1}^N \left[ \frac{\Gamma (2s)}{\Gamma (s - i x_j) \Gamma (s + i x_j)} \right]^{2N}
\\
&\times
\prod_{1 \leq j \leq k \leq N} \Gamma (i x_j + i x_k) \Gamma (- i x_j - i x_k) \Gamma (i x_j - i x_k) \Gamma (- i x_j + i x_k)
\, .
\nonumber
\end{align}

This completes the solution of the open spin chain with soft-hard boundaries in the Separated Variables for the Hamiltonian $\mathcal{H}_{\mathcal J}$ commuting with the top 
off-diagonal $B$-element of the monodromy matrix. As we alluded to above, to find the wave functions $\widetilde\Psi_{p,\bit{\scriptstyle x}}$ for the bottom 
off-diagonal $C$-entry, on has to perform an inversion via Eq.\ \re{InversionOperator},
\begin{align}
\widetilde\Psi_{p,\bit{\scriptstyle x}} = \mathcal{J} \Psi_{p,\bit{\scriptstyle x}}
\, .
\end{align}
The outcome of this operation is shown in the right panel of Fig.\ \ref{WaveFunctionsFig} with the top vertex given in Fig.\ \ref{TopVertexFig}.

\section{Discussion and conclusions}

Having found the complete basis of functions governed by multi-particle dynamics of flux-tube excitations in the presence of soft and hard boundaries, we 
can decompose the subtracted correlation function \re{SubtractedR} as
\begin{align}
\mathcal{R}^\gamma
=
\vev{\mathcal{O}_{\rm bot} \bar{\mathcal{O}}^\gamma_{\rm top}}_{\rm subtracted}
=
\int_0^\infty d p \, {\rm e}^{i p \gamma}
\int \prod_{j=1}^{N-1} \frac{d x_j}{2 \pi} \mu^{-1} (\bit{x}) \vev{\mathcal{O}_{\rm bot} | \widetilde\Psi_{p,\bit{\scriptstyle x}} } \vev{\widetilde\Psi_{p,\bit{\scriptstyle x}} | \bar{\mathcal{O}}_{\rm top} }
\, ,
\end{align}
where $\gamma$ encodes the shifted conformal frame for the top operator with respect to the bottom. It arises in higher than eight correlation function in two-dimensional
kinematics, with the reciprocal variable being the recoil momentum of the soft boundary. The separated variables $\bit{x} = (x_1, \dots, x_{N})$ play the role of the 
flux-tube excitations' rapidities. Upon proper interpretation, this expansion is akin to the pentagon expansion of the Wilson loop on null polygonal contours \cite{Basso:2013vsa} that was 
analyzed in terms of the Separated Variables in Ref.\ \cite{Belitsky:2014rba}. 

Our present consideration can be generalized in a straightforward fashion to a situation when the soft boundary possesses the value of the conformal spin different from the ones of particles in 
the chain interior. The integrable system in this case is an inhomogeneous open spin chain. A first step in this direction was undertaken in Ref.\ \cite{Braun:2018fiz}. One can equally consider 
a kinematical situation when both boundaries become soft and therefore dynamical. This case was analyzed a couple of decades ago within the context of QCD within the framework of 
high-twist quark-gluon-quark operators, when the flux-tube is sourced by fundamental matter fields with gluons propagating in the middle \cite{Braun:1998id,Belitsky:1999qh,Derkachov:1999ze}.

Possibly, the partial light-cone limit considered in this paper could provide a bridge between the pentagon and hexagon frameworks alluded to above for nonperturbative calculation of  
amplitudes and correlators, respectively. This calls for a detailed consideration of how much of the current one-loop analysis can be bootstrapped to all orders in 't Hooft coupling. For the 
correlation function studied in this work, the factorization of the front and back into independent observables is violated at higher orders of the perturbative series, the two faces start interacting 
in spite of the devised subtraction. 

However, the sought after connection between hexagons and octagons can be studied in more basic observables like three (four) point correlation 
functions of two (three) BPS and one spin-$S$ twist-$L$ Wilson operator from the SL(2) sector. As one increases the number of magnons $S \to \infty$, one anticipates emergence of the 
flux-tube, while inclusion of $L-2$ holes introduces rapidities of corresponding flux-tube excitations, and thus would provide an explicit relation between the two formalisms \cite{BasBelXX}. 
This is particularly encouraging in light of the recent discovery that the same (octagon) anomalous dimension \cite{Belitsky:2019fan} governs the Sudakov-like asymptotics of the null limit of 
four-point correlators of infinitely-charged BPS operators \cite{Coronado:2018ypq}, on the one hand, and the behavior at the origin of the six-point gluon scattering amplitude \cite{Basso:2020xts}, 
on the other.

\appendix
\section{Appendix}
\label{FeynmanRulesAppendix}

In this appendix, we summarize the main rules in handling rungs in two-dimensional Feynman graphs, which are indispensable in various calculations in the body of the
paper. Their proof can be found in the literature, see, e.g., \cite{Derkachov:2002tf,Derkachov:2003qb,Belitsky:2014rba}. 

\begin{itemize}

\item Chain rule:
\begin{figure}[h]
\begin{center}
\mbox{
\begin{picture}(0,30)(130,0)
\put(0,-390){\insertfig{20}{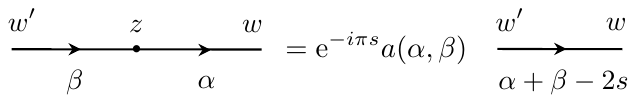}}
\end{picture}
}
\end{center}
\end{figure}

\noindent where
\begin{align}
\label{aChain}
a (\alpha, \beta) = \frac{\Gamma (\alpha + \beta - 2 s) \Gamma (2s)}{\Gamma (\alpha) \Gamma (\beta)}
\, .
\end{align}

\item Cross relation:
\begin{figure}[h]
\begin{center}
\mbox{
\begin{picture}(0,60)(225,0)
\put(0,-355){\insertfig{20}{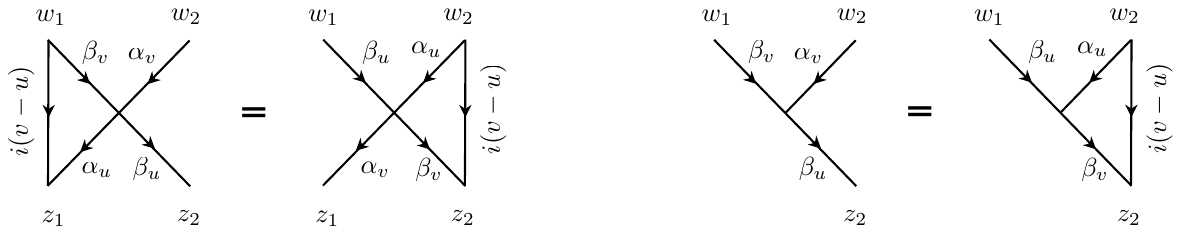}}
\end{picture}
}
\end{center}
\end{figure}

\item Fourier transform:
\begin{align}
\label{FourierTransform}
\int D_s w \frac{{\rm e}^{i p w}}{(z - \bar{w})^\alpha} = \frac{\Gamma (2s)}{\Gamma (\alpha)} p^{\alpha - 2 s} {\rm e}^{-i \pi \alpha/2} {\rm e}^{i p z}
\, ,
\end{align}
for $p>0$.

\end{itemize}


\end{document}